\documentclass[pdflatex,sn-basic]{sn-jnl}

\usepackage{comment}
\usepackage{graphicx}%
\usepackage{multirow}%
\usepackage{amsmath,amssymb,amsfonts}%
\usepackage{amsthm}%
\usepackage{mathrsfs}%
\usepackage[title]{appendix}%
\usepackage{xcolor}%
\usepackage{textcomp}%
\usepackage{manyfoot}%
\usepackage{booktabs}%
\usepackage{algorithm}%
\usepackage{algorithmicx}%
\usepackage{algpseudocode}%
\usepackage{listings}%
\usepackage{tikz}
\usetikzlibrary{decorations.pathmorphing,arrows.meta,positioning,calc,patterns}

\theoremstyle{thmstyleone}%
\theoremstyle{thmstyletwo}%

\theoremstyle{thmstylethree}%

\begin{document}
\title[Cosmology as Representation]{Cosmology as Representation: Informational Invariance and the Limits of Scientific Realism}

\author[1,2]{Stefano Profumo}%
 \email{profumo@ucsc.edu}
\affil[1]{Department of Physics, 
University of California Santa Cruz}
\affil[2]{Santa Cruz Institute for Particle Physics,
Santa Cruz, CA 95064, USA}

\abstract{
Modern cosmology is often taken to provide an increasingly accurate description of the universe’s underlying ontology through progressively refined mathematical models. I challenge this interpretation by arguing that the empirical success of cosmology underdetermines not only ontology but also the mathematical and conceptual frameworks used to represent observational data. I propose instead that the objective content of cosmology is best identified with cross-representational informational invariants—features of observational structure that persist across empirically adequate descriptions. These invariants can be characterized in information-theoretic terms, including correlation structure, statistical distinguishability, and limits on accessible information, and formalized using tools such as the Fisher–Rao metric on model space. On this view, cosmological models are best understood as efficient encodings of observational structure rather than uniquely privileged descriptions of fundamental reality. Scientific progress, accordingly, consists not in convergence toward a fixed ontology, but in the progressive refinement of the informational structure accessible to observation.
}


\maketitle

\section{Introduction}

What does modern cosmology tell us about the universe? A natural answer is
that it reveals, with increasing precision, the large-scale structure and
composition of reality: a spacetime geometry governed by general relativity
and populated, dominantly, by baryons, cold dark matter, and dark energy.
The empirical success of the standard cosmological model, $\Lambda$CDM,
strongly encourages this interpretation. Across a wide range of independent
probes---including the cosmic microwave background, large-scale structure,
gravitational lensing, baryon acoustic oscillations, and Type~Ia
supernovae---a relatively compact parameterization provides both an
extraordinarily accurate fit to observational data and a large number of
falsifiable predictions~\citep{Planck:2018,Weinberg2013,Percival2010,Riess2016}.

It is therefore natural to regard $\Lambda$CDM as a paradigmatic case of
successful scientific representation: a framework in which mathematical
structure, empirical adequacy, and physical interpretation appear to align.
Yet this very success gives rise to a tension. The empirical adequacy of the
model is exceptionally strong, but the ontological conclusions often drawn
from it are less secure. The usual inference moves from predictive success to
ontological commitment---from the successful fitting of correlation functions,
distance measures, and growth histories to the claim that the universe is in
fact composed of the particular entities or components posited by the
model. The legitimacy of that inference is not self-evident.

This paper argues for a more cautious interpretation of cosmological
success. My central claim is that cosmology underdetermines not only the
ontology of the universe, but also, in important cases, the mathematical and
conceptual frameworks through which observational data are represented. If
this is right, then the empirically secured content of cosmology should not be
identified straightforwardly with the ontology of any one successful model.
Instead, it is better located in features of observational structure that
remain stable across empirically adequate representations.

The proposal developed here is that these robust features are best understood
as \emph{informational invariants}: patterns of correlation, statistical
distinguishability, and constrained accessibility that persist across
representational choices. On this view, cosmological models such as
$\Lambda$CDM are best understood not as uniquely privileged descriptions of
fundamental reality, but as highly successful schemes for encoding
observational structure. The aim is not to deny that cosmology yields
objective knowledge, nor to collapse into instrumentalism, but to identify a
more defensible locus of objectivity than theory-internal ontology.

The motivation for this shift is cumulative. First, cosmology inherits the
familiar problem of underdetermination: empirical success does not by itself
fix the existence or nature of the unobservable entities a theory
posits~\citep{vanFraassen:1980,Laudan:1981,Stanford:2006}. Second,
cosmological data are not simply given; they are extracted through
instrumental modeling, foreground subtraction, calibration procedures, and
statistical inference, so that the evidential base is theory-laden in a
substantive sense~\citep{Kuhn:1962,vanFraassen:1980}. Third, modern physics
offers multiple reasons to resist the assumption that successful
representation must be unique. Distinct formulations of quantum mechanics,
cases of duality in high-energy theory, and contemporary debates about the
status of spacetime and subsystems all suggest that empirical content may be
preserved across formally distinct frameworks~\citep{Dawid:2013,Weatherall:2019a}.
Taken together, these considerations weaken the inference from empirical
success to uniquely determined ontology.

What is new here is not the general thought that structure matters more than
objects, nor the familiar claim that cosmology involves substantial
underdetermination. The distinctive proposal is to locate the relevant
structure \emph{across} empirically adequate representations rather than
\emph{within} any single successful theory, and to characterize that
cross-representational content in informational terms. In this respect, the
view differs from standard structural realism, which typically reads
structure off the formalism of our best current theories, and from
empiricism, which treats the relevant constraints as epistemic rather than as
candidates for objectively fixed content~\citep{Worrall:1989,Ladyman:1998,French:2014,vanFraassen:1980}.
The claim is not that ontology is irrelevant, but that the most secure
achievements of cosmology lie at the level of informational structure that
survives representational variation.

To make this proposal more precise, I draw on information geometry.
In particular, the Fisher--Rao metric on model space provides a natural way
to formalize statistical distinguishability independently of arbitrary
parameterization. By \v{C}encov's theorem, it is the unique Riemannian metric
(up to overall scaling) invariant under sufficient
statistics~\citep{Cencov:1982}. This does not, by itself, settle any
metaphysical question. But it does supply an operational framework for
distinguishing representational artifacts from features of model space that
are invariant under admissible transformations, and thus for clarifying what
it would mean to treat certain informational features as objectively
significant. Recent work has shown the fruitfulness of such tools in
cosmological inference~\citep{Giesel:2021}.

The broader philosophical consequence is a revised form of realism about
cosmology. Against constructive empiricism, I argue that cosmology does
track objective features of the world; against more standard realist
readings, I argue that these features are not best understood as the
specific ontological posits of any one model~\citep{vanFraassen:1980,Psillos:1999,Putnam:1975,Callender2017}.
The position developed here---which I call \emph{informational structural
realism}---holds that what cosmology secures most robustly is invariant
informational structure. Scientific progress, on this account, consists less
in convergence toward a uniquely determined ontology than in the progressive
refinement of the informational structure that observation can constrain.

My aim is therefore not to diminish the achievements of cosmology, but to
clarify their epistemic significance. If the argument succeeds, cosmology
does not straightforwardly tell us what the universe is made of. Rather, it
identifies increasingly refined constraints on the informational structure of
the observable universe, constraints that any empirically adequate
representation must respect.

The paper proceeds as follows. Section~\ref{sec:underdetermination} examines underdetermination in cosmology and the assumptions about scientific representation that encourage ontological overreach. Section~\ref{sec:pressure} argues that developments in modern physics and the inferential character of cosmological observation undermine the assumption that successful representation must be unique. Section~\ref{sec:invariants} introduces informational invariants as the appropriate locus of objective content and develops their characterization in terms of correlation structure, distinguishability, and limits on accessible information. Section~\ref{sec:ISR} situates informational structural realism relative to ontic structural realism and constructive empiricism, clarifying its distinctive commitments. Section~\ref{sec:info_geom} develops the formal role of information geometry, showing how the Fisher--Rao metric and Kullback--Leibler divergence provide an operational account of invariance and equivalence classes in cosmological inference. Section~\ref{sec:objections} addresses objections and limitations, including concerns about empiricism, observer-relativity, admissibility, and the scope of the formal framework. Section~\ref{sec:conclusion} concludes by drawing out the implications of this view for the interpretation of cosmological knowledge and for the nature of scientific progress.

\section{Underdetermination, the Galilean Program, and Ontological Modesty}
\label{sec:underdetermination}

The tension identified in the Introduction can be stated more precisely in
terms of underdetermination. The general point is familiar: empirical
success does not, by itself, uniquely fix the ontology of a theory. What
matters for present purposes is not merely that this problem exists in the
abstract, but that it arises repeatedly in the history of physics, is
reinforced by the methodological situation of cosmology, and bears directly
on the interpretation of contemporary cosmological models.


Several familiar episodes in the history of physics illustrate the gap
between empirical adequacy and ontological commitment. The luminiferous
ether was once regarded as indispensable for explaining electromagnetic wave
propagation, yet its eventual abandonment did not occur because the theory
had ceased to organize phenomena effectively; rather, alternative
formulations made the ontological posit dispensable. Caloric theory likewise
accounted successfully for a wide range of thermodynamic phenomena, and its
replacement by kinetic theory involved not simply improved prediction but a
reorganization of ontology~\citep{Kuhn:1962}. Even Newtonian gravity,
which remains empirically successful over a broad domain, is tied to an
ontology of absolute space, absolute time, and action at a distance that was
not retained in general relativity.

The point of rehearsing these examples is not that successful theories are
usually false, but that success at the level of prediction and explanation
does not by itself settle what kinds of things the world fundamentally
contains. A theory may be empirically powerful and explanatorily unified
while nevertheless misdescribing, or only partially capturing, the ontology
associated with its own formalism.


It is useful, accordingly, to distinguish three levels at which a scientific
theory may succeed~\citep{vanFraassen:1980}:
\begin{enumerate}
\item \emph{Empirical adequacy}: the theory correctly accounts for observable
phenomena and yields accurate predictions.
\item \emph{Explanatory coherence}: the theory organizes diverse phenomena
within a unified and systematically articulated framework.
\item \emph{Ontological commitment}: the entities and structures posited by
the theory are taken to correspond to the fundamental constituents of
reality.
\end{enumerate}

These levels are often run together in scientific practice, especially when
a theory is exceptionally successful. But the historical episodes just noted
show that success at the first two levels does not automatically secure the
third. This distinction will be central below, because much of the force of
standard realist readings of cosmology depends on sliding from empirical and
explanatory success to ontological entitlement.

\subsection{The Galilean Program}
\label{sec:galilean}

That slide is encouraged by a powerful and often implicit picture of what
science is doing. In a compressed form, one may call it the \emph{Galilean
program}: the idea, traceable to Galileo's claim that the book of nature is
written in the language of mathematics, that the aim of science is to
recover the mathematical structure of the world
itself~\citep{Galileo:1623,Drake:1957,Worrall:1989}. In its strongest form,
this picture combines three commitments: that the world possesses a
determinate observer-independent structure, that scientific progress consists
in progressively recovering that structure, and that the empirical success
of a theory provides evidence that its formalism captures the ontology of
the world~\citep{Wigner1960,Psillos:1999}.

This program has been extraordinarily fruitful. Physics has repeatedly
produced mathematically precise frameworks that unify disparate phenomena and
generate strikingly accurate predictions. In cosmology, the appeal of the
program is especially strong. $\Lambda$CDM offers a compact parameterization
that successfully accounts for observations from the cosmic microwave
background to the large-scale distribution of
galaxies~\citep{Planck:2018}, and it is therefore natural to interpret its
central posits---general-relativistic spacetime, cold dark matter, and a
cosmological constant---as corresponding to real constituents of the
universe~\citep{Martens:2022,Kosso:2013}.

The philosophical difficulty is not with the methodological power of this
picture, but with a further assumption often attached to it: namely, that a
successful mathematical representation is, in the relevant sense,
\emph{uniquely privileged}. That assumption is closely connected to the
realist thought that the predictive success of science would be miraculous
if our best theories were not at least approximately true~\citep{Putnam:1975,Musgrave:1988}.
But once the possibility of multiple observationally indistinguishable representations is
taken seriously, the inference becomes less secure. If formally distinct
frameworks can capture the same empirical
content~\citep{Duhem:1954,Quine:1975}, then the success of any one of them
does not by itself establish the ontological correctness of its own
particular posits~\citep{vanFraassen:1980,Laudan:1981}. The historical
episodes just noted already counsel caution on this point~\citep{Kuhn:1962,Stanford:2006}.

I do not mean to reject the Galilean program wholesale. The issue is narrower
but decisive: whether empirical success warrants the stronger claim that one
particular successful representation has thereby earned uniquely privileged
ontological status.

\subsection{Underdetermination in Cosmology}

This question is especially pressing in cosmology. As \citet{Smeenk2014} has
emphasized, cosmology studies a unique system from an embedded observational
position, and therefore lacks many of the strategies that elsewhere help
reduce underdetermination, such as varying initial conditions, manipulating
systems, or repeating experiments under controlled
circumstances~\citep{Smeenk2014,Smeenk2020}. The resulting limitation is not
merely practical. It shapes the inferential situation of the discipline
itself.

This feature amplifies a more general problem. In cosmology, what is often
best established is a pattern in the data, while the ontology invoked to
explain that pattern remains less securely fixed. Cold dark matter, for
example, provides a highly successful representation of gravitational
phenomena across multiple scales, but what the evidence directly fixes is a
persistent discrepancy between visible matter and gravitational effects; the
further interpretation of that discrepancy in terms of a particular particle
species is not uniquely forced by the data. Likewise, the cosmological
constant provides an effective description of the observed late-time
acceleration of cosmic expansion, but whether it should be understood as
vacuum energy, as an effective parameter, or as a sign of deeper new
structure remains unsettled. Inflation offers an elegant account of the
origin of large-scale structure and of the statistical properties of
primordial perturbations, yet a wide range of inflationary constructions can
reproduce the same broad observational signatures.

The same pattern recurs in each case. Empirical adequacy is substantial.
Explanatory coherence is often impressive. But the ontological conclusions
that are frequently drawn from these successes remain underdetermined. In
cosmology, that underdetermination is not an occasional anomaly; it is built
into the epistemic situation of the field.

\subsection{Ontological Modesty and the Turn to Structure}

These considerations motivate a stance of {\it ontological modesty}. This is not a
retreat to instrumentalism. The empirical success of cosmology is too
systematic, too resilient across probes, and too theoretically productive to
be dismissed as a mere calculating device. It is entirely reasonable to
maintain that cosmological theories track genuine features of the world. The
question is which features they track most securely.

A natural answer is provided by structural realism
\citep{Worrall:1989,Ladyman:1998,French:2014}. On that view, what survives
theory change most robustly is not necessarily a specific inventory of
unobservable entities, but the structure encoded in successful theories. I
am sympathetic to that line of thought, including its information-theoretic
development in \citet{LadymanRoss:2007}. But the pressure of cosmological
underdetermination suggests a further step. In the present context, the most
robust content is not simply the relational structure internal to a single
successful theory. It is, rather, the content that persists across distinct
but empirically adequate representations. This shifts the relevant locus of
objectivity from structure \emph{within} one formulation to structure
preserved \emph{across} formulations---a move continuous with recent work on
theoretical equivalence in philosophy of
physics~\citep{Weatherall:2019a,Weatherall:2019b}.

If empirical success underdetermines ontology, and if successful
representation need not be unique, then the most defensible candidates for
representation-independent content must lie at a level preserved across representational
variation. That content, I will argue, is best characterized in
informational terms: as the statistical and inferential structure encoded in
observational data, the pattern of dependencies and constraints that any
adequate cosmological model must reproduce~\citep{Floridi:2008,Dennett:1991}.
Underdetermination, on this view, does not merely limit how far ontological
inference may extend. It positively redirects philosophical attention toward
a more stable level of description.


\section{Pressure from Modern Physics}
\label{sec:pressure}

The preceding section argued that empirical success does not by itself
license strong ontological conclusions. A further source of pressure comes
from modern physics itself. The issue is not merely that some contemporary
theories are conceptually unfamiliar or technically difficult. It is,
rather, that modern physics repeatedly exhibits cases in which empirical
content is preserved across formally or conceptually distinct
representations. This weakens the assumption, central to more ambitious
realist readings of cosmology, that a successful theory thereby provides a
uniquely privileged mathematical description of reality.

\subsection{Quantum Theory and the Instability of Classical Categories}

The classical picture presupposed by much scientific realism is familiar:
the world is composed of well-defined objects bearing intrinsic properties,
evolving in a spacetime arena according to deterministic laws
\citep{Newton:1687,Laplace:1814,Earman:1986}. Composite systems are taken to
be analyzable into independently characterizable parts, and physical
properties are assumed to be well defined independently of acts of
measurement. Cosmological models are often interpreted against this
background. On such a reading, the universe has a definite physical state,
its metric structure evolves dynamically, and its contents---baryons, dark
matter, dark energy, radiation---are naturally treated as entities or fields
inhabiting spacetime~\citep{Wald:1984}.

Quantum theory places this picture under strain. Most obviously,
entanglement undermines separability: the state of a composite system cannot,
in general, be decomposed into independently well-defined states of its
parts~\citep{Schrodinger:1935,Bell:1964}. This is not merely an epistemic
limitation but a structural feature of the theory, confirmed experimentally
\citep{Aspect:1982,Hensen:2015}, and it puts pressure on the idea that the
world is fundamentally composed of localized objects with intrinsic
properties~\citep{Howard:1985,Teller:1986}. The measurement problem deepens
the difficulty by making it unclear in what sense physical properties may be
regarded as determinate prior to observation~\citep{Bell:1987,Maudlin:1995}.

For present purposes, however, the crucial point is not any one proposed
resolution of these difficulties. It is that distinct interpretations of
quantum mechanics---including Everettian, Bohmian, and collapse-theoretic
approaches---offer sharply different ontological pictures while preserving
the same observable predictions~\citep{Everett:1957,Bohm:1952,GRW:1986,Albert:1992}.
Even where a common formal core is shared, ontology is not uniquely fixed by
empirical success. That fact alone is enough to undermine the idea that a
successful physical theory must wear its ontology transparently on its
formal sleeve.

The lesson for the present argument is limited but important. Quantum theory
does not merely add novel entities to an otherwise stable conceptual
framework; it shows that categories such as object, separability, and
pre-observational determinacy cannot simply be assumed to retain their
classical roles across all successful physical descriptions
\citep{Bohr:1935,Heisenberg:1958}. If ontology is not uniquely read off even
from one of the most successful theories in physics, there is reason for
caution when drawing strong ontological conclusions from cosmological
success.

\subsection{Representational Non-Uniqueness}

Modern physics also provides more direct evidence that the same physical
content may be encoded in more than one mathematical framework. The point
does not depend on any one exotic example; it appears in both familiar and
more speculative settings.

Within quantum theory itself, the Schr\"odinger, Heisenberg, and path-integral
formulations differ significantly in mathematical structure and conceptual
presentation, yet are standardly understood to be equivalent formulations of
the same theory~\citep{Dirac:1930,Feynman:1948,vonNeumann:1932}. Their
historical unification already established an important philosophical point:
physical content need not be tied to one uniquely preferred mathematical
realization~\citep{Muller:1997a,Muller:1997b,Perovic:2008}.

A more dramatic example is provided by duality. To take one prominent case,
the AdS/CFT correspondence relates a gravitational theory in
$(d{+}1)$-dimensional anti-de Sitter spacetime to a conformal field theory
without gravity on its $d$-dimensional boundary
\citep{Maldacena:1999,Witten:1998}. The two descriptions differ strikingly
in ontology and mathematical presentation, yet are taken to encode the same
physical content. Whatever one ultimately thinks about the metaphysical
status of such dualities, they make it difficult to sustain the view that
successful representation must always be unique, or that ontology can be
straightforwardly read off from the manifest features of one successful
formalism~\citep{Rickles:2011,Dawid:2013}.

Recent philosophy of physics has made this point more precise.
\citet{ReadMollerNielsen:2020} argue that many dual theories are best
understood not as genuinely competing ontological claims but as alternative
presentations of the same theory. Related work by \citet{DeHaro:2017},
\citet{DeHaroButterfield:2018}, \citet{Menon:2019}, and
\citet{Weatherall:2019a,Weatherall:2019b} shows that questions of
theoretical equivalence and representational redundancy are central to any
responsible inference from formalism to ontology. The general moral is not
that all successful theories are equivalent, but that formal distinctness
does not by itself imply difference in physical content.

This point is reinforced by the broader semantic conception of scientific
theories, on which models function as mediating representational structures
rather than as transparent mirrors of the world
\citep{Giere:1988,vanFraassen:2008,FriggNguyen:2021,Suppe:1977,French:2020}.
Once this is taken seriously, the philosophical question is no longer simply
which theory is true, but which features of a successful representational
practice remain stable across changes of formulation, interpretation, or
modeling framework.

\subsection{Cosmology as an Inferential Enterprise}

The case for caution becomes even stronger in cosmology, because
representational plurality appears not only at the level of theory but also
at the level of evidence. Cosmological data are not raw givens. They are
constructed through chains of calibration, foreground subtraction, noise
modeling, statistical compression, and parameter inference
\citep{Planck:2018}. The angular power spectrum of the cosmic microwave
background, for example, is not directly observed in anything like an
unmediated form; it is extracted through a highly structured inferential
pipeline.

This does not undermine the reliability of cosmological results. On the
contrary, the robustness of those results is often demonstrated by the fact
that independent pipelines, instruments, and probes converge on closely
related constraints. But it does undermine a simpler picture according to
which theory merely reads off the structure of the world from uninterpreted
data. In cosmology, observational content is inseparable from modeling and
inference. That point becomes especially vivid when tensions arise between
different data sets or analysis choices, as in ongoing disputes over the
Hubble parameter. What is exposed in such cases is not the failure of
observation, but the fact that observational outputs are themselves
representation-dependent in ways that require philosophical attention.

The lesson is therefore continuous with the one drawn above from quantum
theory and duality. Even before one reaches ontology, cosmology already
operates through representationally structured access to the world. The
success of a cosmological model cannot simply be identified with a direct
mapping between theory-internal posits and mind-independent reality.

\subsection{Implications}

The examples considered in this section differ substantially from one
another. Empirical equivalence between quantum interpretations, equivalence
between formulations, duality relations in high-energy theory, and
representation-dependence in observational cosmology are not instances of a
single formal relation. The present argument does not require assimilating
them into one notion of equivalence. What matters is the common lesson they
support: empirical content may be preserved across distinct mathematical,
conceptual, or inferential representations, so the inference from the
particular features of one representation to ontological conclusions is less
secure than standard realist readings often assume.

This does not show that ontology is irrelevant, nor that all representations
are equally good. It shows, rather, that the success of a representation
does not automatically confer uniquely privileged ontological status on its
own internal categories. These cases therefore motivate locating physical
content at the level of what is preserved across representations. In the
cosmological context, I will argue, those preserved features are best
understood in informational terms.


\section{Toward Informational Invariants}
\label{sec:invariants}

If cosmology underdetermines not only ontology but also, in some cases, the
mathematical and conceptual frameworks used to represent observational data,
a natural question follows: what, if anything, remains fixed? What does
cosmology secure, if not the uniquely correct ontology of a single successful
model?

Informational structural realism (ISR) identifies the epistemically stable
content of cosmology with features that satisfy three criteria:
\begin{enumerate}
\item \emph{Representation-independence}: they are not tied to the
particular ontology, parameterization, or formal machinery of any one
successful model;
\item \emph{Empirical extractability}: they are recoverable, however
theory-mediatedly, from observational practice;
\item \emph{Stability across admissible descriptions}: they persist under
transitions between representations that remain empirically adequate at the
relevant observational resolution.
\end{enumerate}
These criteria identify not objects or substances, but patterns of
constraint: what the data correlate, what the data can distinguish, and what
the observational situation prevents us from accessing.

The point is more specific than the generic claim that ``structure'' matters.
Once representational non-uniqueness is taken seriously, the strongest
candidates for invariant informational structure are those that survive
variation in ontology, formalism, and inferential pipeline. In cosmology,
three families are especially important: correlation structure,
distinguishability structure, and information bounds.

A qualification is needed at the outset. The candidates for informational
invariants in cosmology are not raw, theory-free deliverances of experience.
Power spectra, distance--redshift relations, covariance matrices, and Fisher
eigendirections are all produced through instrument models, likelihood
choices, foreground subtraction, and statistical compression. ISR does not
deny this. What matters is not that an invariant be free of theory, but that
it remain stable across the range of theoretical and inferential choices
compatible with empirical success. The invariance in question is therefore
\emph{cross-representational}, not pre-theoretical.

\subsection{Three Families of Informational Invariants}

\subsubsection{Correlation structure}

The most immediate candidates for informational invariants are correlation
patterns among observables. In cosmology, the empirical content of a model is
often encoded in the statistical structure of observational fields: the
angular power spectrum of the cosmic microwave background, the two-point and
higher-order correlation functions of galaxy clustering, and the matter power
spectrum reconstructed from weak lensing are paradigmatic
examples~\citep{Planck:2018,BOSS:2017,DES:2022,Tegmark:1997,Verde:2010}.
These are not themselves ontological claims, but specifications of
statistical dependence among observationally accessible degrees of freedom.

The CMB provides the clearest illustration. What observation fixes most
directly is not any one theory's preferred microphysical story, but a highly
structured pattern of anisotropies and acoustic peak relations. Those
correlations can be derived within $\Lambda$CDM by evolving primordial
perturbations through coupled Boltzmann and Einstein equations in an
expanding background \citep{Ma:1995,Seljak:1996}, but the observational
content is not exhausted by that derivation. What is secured are the
positions, amplitudes, and relative structure of the peaks, the broad shape
of the spectrum, and the covariance properties of the inferred multipoles.
Different theoretical frameworks may supply different dynamical mechanisms
while remaining constrained to reproduce that same pattern to the relevant
precision.

This is why correlation structure is a natural candidate for invariant
content. Its significance does not depend on whether the underlying ontology
is phrased in terms of cold dark matter particles, effective fluids, modified
gravity, or some more fundamental description. What any empirically adequate
representation must recover is the structured pattern of dependence itself.

\subsubsection{Distinguishability structure}

A second family of invariants concerns not the correlations among observables
directly, but the geometry of \emph{distinguishability} over the space of
models. Cosmological data constrain not only whether a model is broadly
acceptable, but \emph{how well} nearby alternatives can be told apart. That
structure can be formalized by the Fisher information matrix, which equips
model space with a Riemannian metric---the Fisher--Rao metric---whose local
geometry encodes the statistical separability of competing
parameterizations~\citep{Amari:2000,Cencov:1982,Heavens:2009,Giesel:2021,Schaefer:2016}.

The philosophical importance of this move is easy to miss if Fisher methods
are treated merely as technical forecasting tools. They provide a
representation-independent characterization of which directions in model
space are tightly constrained, which are weakly constrained, and which remain
degenerate at a given observational resolution. A degeneracy direction is not
simply a nuisance of notation; it is a structural fact about the
informational landscape. Conversely, sharply constrained eigendirections
identify combinations of parameters or model features that the data resolve
robustly.

Here again the CMB is illustrative. Observations do not typically determine
each theory parameter in isolation. They constrain specific combinations:
baryon density and matter density affect peak structure differently;
reionization, scalar amplitude, and lensing contributions are partially
degenerate; curvature, dark-energy history, and late-time distance measures
interact in ways that depend on the probe and its
precision~\citep{Planck:2018,Tegmark:1997,Heavens:2009}. For ISR, these
patterns of distinguishability are more fundamental than the coordinate
labels attached to a model. Parameter names and parameterizations may vary;
the underlying geometry of what the data can and cannot distinguish is the
more stable content.

Correlations describe the empirical patterns a model must reproduce.
Distinguishability describes the inferential resolution with which competing
models can be separated. Together they specify not only what cosmology
observes, but what it is in a position to know.

\subsubsection{Information bounds and constrained accessibility}

A third family of invariants concerns the limits on extractable information.
Cosmology is unusual among the sciences in that those limits are not merely
contingent experimental inconveniences. They are partly fixed by the causal
and statistical situation of observers embedded within the universe.

One aspect of this limitation is causal accessibility. Cosmological
observation is confined to the observer's past light cone, and more
generally to the horizon structure of the spacetime accessible to signal
exchange~\citep{Ellis:1975,Ellis:2014,Rindler:1956,Penrose:1972,Malament:1977}.
Even if spacetime itself turns out to be non-fundamental, the distinction
between what can and cannot contribute to observation remains a robust
constraint on physical description. Another aspect is statistical:
cosmological inference is subject to cosmic variance, especially on large
angular scales, because observers have access to only one realization of the
relevant stochastic processes~\citep{Knox:1995,Scott:2016}. No improvement in
detector sensitivity can eliminate that limit. A third aspect is more
general and information-theoretic: entropy bounds and related arguments place
upper limits on the information content accessible within bounded regions or
horizon-limited domains~\citep{Bekenstein:1981,Bousso:1999,Bousso:2002}.

These are naturally understood as bounds on \emph{constrained accessibility}.
They specify not merely what has not yet been measured, but what cannot be
made available to embedded observers beyond certain thresholds. Any adequate
representation must respect them. One lesson of modern cosmology is that what
can be known is itself physically structured.

\subsection{Equivalence Classes and the Locus of Objectivity}
\label{sec:equivalence_classes}

Once invariant informational features are relocated from theory-internal
ontology to informational structure, the natural unit of philosophical
analysis is no longer an individual model taken in isolation. It is an
\emph{equivalence class} of models that share the same informational content
at a given level of observational resolution.

The underlying idea has familiar analogues in physics and philosophy.
Gauge-related models in field theory, diffeomorphically related
representations in general relativity, and other cases of theoretical
equivalence all encourage the thought that physically meaningful content is
not exhausted by the particular representative through which it is expressed
\citep{Healey:2007,Earman:1989,Norton:1993,Weatherall:2016,Barrett:2019,Halvorson:2012}.
ISR extends that lesson to cosmology. Models may differ in ontology, formal
structure, or explanatory vocabulary while still belonging to the same
informational class so long as they remain empirically indistinguishable at
the relevant level of precision.

The notion of equivalence at work here should be stated carefully. ISR does
not require strict identity of formalism or exact identity of all logically
possible predictions. What matters is approximate empirical
indistinguishability relative to a specified observational configuration. Two
models belong to the same equivalence class, relative to an observational
setup $O$, if they generate observational distributions that cannot be
distinguished by the data available under $O$ at the required threshold of
precision. In Section~\ref{sec:info_geom}, this idea will be given a more
explicit formal expression using information geometry and
Kullback--Leibler-type measures of separation. For present purposes, the key
point is conceptual: the boundary between classes is set by observational
resolution, not by a priori metaphysical preference for one ontology over
another.

This has an important consequence for how scientific progress should be
understood. On a standard realist picture, progress is often described as
convergence toward the true ontology or toward the correct values of the
parameters appearing in the true theory. ISR suggests a different picture.
As observational precision improves, equivalence classes split. Directions in
model space that were once degenerate become distinguishable. Correlation
patterns once only coarsely measured become more sharply resolved. What
increases is not necessarily proximity to a uniquely fixed ontology, but the
resolution of the informational structure accessible to observation.

\subsection{Admissibility and Domain-Relativity}

A prior question arises at this point: which representations count as
eligible members of an equivalence class? If the admissible class is left
completely unconstrained, the framework risks triviality. If it is defined
too narrowly, the proposal collapses back into the assessment of notational
variants within one favored theory.

The minimal answer adopted here is deliberately permissive. A representation
is admissible if it (i) addresses the same observational domain, in the
sense that it generates probability distributions over a common data space;
(ii) is mathematically well defined enough to yield a computable or at least
well-posed likelihood structure; and (iii) is not already excluded by
existing data at the relevant level of precision. These criteria admit a
fairly broad range of cosmological representations while still preventing the
equivalence relation from becoming vacuous.

This also shows why the resulting invariants are \emph{domain-relative}
without thereby being subjective. The informational invariants of CMB
cosmology are not identical to those of large-scale structure or supernova
cosmology, even though formal tools may overlap. Objectivity here is always
indexed to a definite empirical domain. That is not a weakness of the view.
ISR is not offering a theory-independent inventory of everything real; it is
offering an account of what, within a given observational context, is most
defensibly taken as representation-independent content.

\subsection{Observation, Configuration, and Embedded Access}

A further nuance is crucial. Equivalence classes are defined not only
relative to an observational domain but also relative to an
\emph{observational configuration}: sky coverage, noise properties,
instrumental bandpass, angular resolution, redshift reach, foreground model,
and the like. This dependence is not an embarrassment to the framework; it
is one of its central insights.

The Fisher geometry of model space, for example, depends on the observing
setup as well as on the model family itself~\citep{Tegmark:1997,Heavens:2009,Schaefer:2016}.
Two models that are empirically indistinguishable with Planck-like noise and
resolution may become distinguishable for a more sensitive future CMB
experiment. What changes in such cases is not the retrospective truth value
of the earlier analysis, but the partition of theory space induced by new
informational access. The observational situation does not merely reveal a
pre-existing ontology more clearly; it partly determines which differences
between representations are empirically meaningful.

This is particularly important in cosmology because observers are embedded
within the universe they study. The available informational structure is
shaped by causal horizons, finite mode counts, and cosmic variance. ISR
therefore does not seek a ``view from nowhere.'' It seeks an account of
objectivity appropriate to observers situated within a physically structured
epistemic horizon. What counts as invariant is what remains stable across
admissible representations \emph{for such observers}, given the access they
can in principle have.

\subsection{Transition}

The proposal can now be stated compactly. Cosmology secures not primarily a
uniquely correct ontology, but cross-representational informational
invariants: stable patterns of correlation, stable structures of
distinguishability, and stable bounds on accessible information. These
invariants are defined relative to admissible representations and
observational configurations, but they are not thereby merely subjective or
conventional. They are the most robust features of what the world makes
available to cosmological inquiry.


\section{Informational Structural Realism Defined and Distinguished}
\label{sec:ISR}

In what follows, I clarify how informational structural realism
differs from its nearest philosophical neighbors, and why those differences
matter in the specific epistemic setting of cosmology. 
The contrast may be stated in deliberately schematic form:\\
\begin{quote}
\begin{tabular}{ll}
\textbf{View} & \textbf{What is taken as real} \\
\hline
Ontic structural realism & Structure within a successful theory \\
Constructive empiricism & Observables, or empirical adequacy \\
Informational structural realism & Cross-representational informational invariants
\end{tabular}
\end{quote}

\vspace*{0.5cm}
The table is not meant to replace the more careful discussion that follows,
but it highlights the central difference in orientation. ISR agrees with
structural realism that scientific success secures more than a merely
instrumental relation to the world, and with empiricism that ontology is
underdetermined by empirical success. Its distinctive claim is that the most
defensible target of realist commitment lies neither in theory-internal
ontology nor in the observable alone, but in the informational structure that
remains stable across admissible representations.

\subsection{Ontic Structural Realism}

The most developed form of structural realism is the \emph{ontic} variant due
to Ladyman, Ross, and French~\citep{LadymanRoss:2007,French:2014}. On this
view, the world is fundamentally structural: what exists are relations and
patterns rather than objects with intrinsic properties. The appeal of this
position is clear. It preserves the realist thought that science captures
objective features of the world while avoiding commitment to the full
ontology of any theory whose object-language may later be revised. Its
historical motivation is equally familiar: across theory change, it is often
relational and mathematical structure, rather than specific unobservable
objects, that appears most stable~\citep{Worrall:1989}.

ISR is sympathetic to the impulse behind OSR. In cosmology, where the
temptation to move too quickly from empirical success to ontological
commitment is especially strong, any view that counsels restraint while
preserving realism is already an advance over naive scientific realism. The
difficulty, however, is that OSR still tends to identify the relevant
structure with the mathematical content of our best current theory. The
structures taken to be real are those encoded in the formalism of quantum
field theory, general relativity, or whatever theory is given privileged
status at the time: state spaces, symmetry groups, field equations, and the
like~\citep{LadymanRoss:2007}.

That move is less secure in contemporary cosmology and fundamental physics
than it may first appear. It presupposes that the successful theory in
question occupies a uniquely privileged representational position. If a
single formalism is the right one, then reading structure off that formalism
is reasonable. But the preceding sections have argued that this
presupposition is precisely what is under pressure. Dualities, equivalent
formulations, theory-laden inference, and framework-dependent observational
outputs all weaken the idea that mathematical success singles out one
uniquely adequate representational scheme. Once that pressure is taken
seriously, the structure \emph{within} any one formalism can no longer be
straightforwardly identified with the structure \emph{of} the world.

\subsection{Three Points of Departure from Ontic Structural Realism}
\label{sec:departures}

Informational structural realism departs from OSR in three connected respects.
Each responds to a genuine difficulty in the more familiar structural realist
picture, and each is motivated by the epistemic and methodological situation
of cosmology itself.

\subsubsection{Structure across representations, not within one}
\label{sec:across}

The first and most fundamental difference concerns where the relevant
structure is located. For OSR, structure is typically read off the formalism
of a particular theory: the permutation symmetry of quantum mechanics, the
gauge structure of the Standard Model, or the diffeomorphism invariance of
general relativity~\citep{French:2014,Ladyman:1998,French:2006}. On that
approach, the world's structural content is identified with the mathematical
relations encoded in the equations of our best current theory. ISR relocates
the issue. The structure that carries the strongest claim to objectivity is
not the structure internal to any one successful formalism, but the structure
preserved \emph{across} empirically adequate representations.

This shift is not verbal. It addresses a persistent problem for structural
realism: how to distinguish physically significant structure from
representational artifact \citep{Frigg:2011,Psillos:2006,Ainsworth:2010}. If
one begins with a single formalism, the entire burden falls on an internal
discrimination: which parts of the mathematics are world-involving, and which
are merely scaffolding? ISR changes the starting point. A feature counts as a
serious candidate for empirically firm content only if it survives
transitions between distinct descriptions that remain empirically adequate.
The burden of realism is thus shifted from reifying parts of one theory to
identifying what different successful theories cannot help but share.

The angular power spectrum of the cosmic microwave background provides a
useful illustration. Within $\Lambda$CDM, the observed $C_\ell$ spectrum is
derived from a specific set of coupled Boltzmann and Einstein equations
governing perturbations in a Friedmann--Lema\^itre--Robertson--Walker
background~\citep{Ma:1995,Seljak:1996}. Within a modified-gravity framework,
such as a scalar-tensor theory in the Horndeski class~\citep{Horndeski:1974,Bellini:2014},
an empirically comparable spectrum may be generated by a different dynamical
mechanism and a different field content. What ISR treats as objectively real
is not the particular dynamical story told by either framework, but the
correlation structure that both must reproduce. The invariant content lies in
the acoustic peak pattern, the covariance structure, and the observational
constraints they encode, not in the machinery by which one favored formalism
happens to generate them.

More precisely, the relevant structure is preserved under the equivalence
relation defined by shared empirical adequacy at a given observational
resolution (\S\ref{sec:equivalence_classes}). This shifts the locus of
ontological commitment from the interior of a single theory to the
\emph{intersection} of an equivalence class: not what one model says in its
own terms, but what all successful models say in common. In this respect ISR
may be described as a form of second-order structural realism. It is realist
about the structure of the space of admissible
representations---about what remains stable under representational
variation---rather than about the structure encoded in any one representation
taken in isolation.

This move is also continuous with recent work on theoretical equivalence.
Weatherall's framework asks when two theories count as the same theory under
structure-preserving mappings~\citep{Weatherall:2019a,Weatherall:2019b}. ISR
asks a different, though related, question: what invariant informational
structure is shared by theories that need not be identical in that strong
sense, but are empirically indistinguishable at a given level of precision?
The shared lesson is that objectivity is not most securely located in what
varies from formulation to formulation, but in what survives those
variations~\citep{Coffey:2014,Dewar:2019}.

\subsubsection{An operational criterion for invariance}
\label{sec:operational}

A second point of departure concerns precision. One persistent objection to
structural realism is that the notion of ``structure'' is too permissive, too
indeterminate, or too formal to carry the philosophical weight placed upon it
\citep{Psillos:2006,Cao:2003,Newman:1928,Frigg:2011}. If every mathematical
feature of a formalism counts as structure, the view risks triviality. If
only some do, a criterion is needed to say which ones matter and why.

ISR responds by grounding invariance not in the abstract formal properties of
a single theory, but in information-theoretic and inferential features that
can be specified independently of any one ontology. As a reminder, ISR posits that three classes of
structure are central:
\begin{itemize}
\item \emph{Correlation}: the statistical dependencies that any adequate theory must
reproduce;
\item \emph{Distinguishability}: the inferential geometry that determines which model
differences data can and cannot resolve;
\item \emph{Constrained accessibility}: the bounds on extractable information imposed
by horizons, finite mode counts, and cosmic variance.
\end{itemize}
Each admits formal expression. Correlation structure is captured by $n$-point
functions and their generalizations~\citep{Peebles:1980,Bernardeau:2002}.
Distinguishability is formalized by the Fisher information metric and the
associated geometry of model space~\citep{Amari:2000,Cencov:1982,Ly:2017}.
Information bounds are characterized by entropy, mutual information, and
related measures~\citep{Cover:1991,Shannon:1948}. This gives ISR an
operational criterion that OSR lacks: what is objectively significant is what
remains invariant under admissible transformations of representation and
inference.

The importance of this criterion is not merely methodological. It provides a
basis for distinguishing between what the data secure and what a particular
framework adds. A feature counts as invariant if it is preserved under
reparameterization and model transformation as characterized by the relevant
information geometry. Degeneracy directions in Fisher space, for example, are
not defects of a parameterization; they are stable features of the
informational landscape, recording genuine limits on what the data can
distinguish~\citep{Giesel:2021,Schaefer:2016}. By contrast, features that
depend on a coordinate choice in model space, or on one ontology-specific way
of labeling the same constraints, are representation-dependent. The contrast
maps naturally onto the distinction between coordinate-dependent quantities
and geometric invariants in Riemannian geometry~\citep{Amari:2000,Kass:1989}.

The dark matter case makes the force of this distinction especially clear.
When one asks whether dark matter is ``real,'' ISR does not offer either a
full ontological endorsement or a skeptical dismissal. Instead, it separates
two claims often run together. The first is that there exists a highly stable
pattern of gravitational discrepancies, encoded across independent
observational probes: galaxy rotation curves~\citep{Rubin:1980}, the Bullet
Cluster separation of luminous and gravitating mass~\citep{Clowe:2006}, the
shape of the matter power spectrum~\citep{BOSS:2017}, and the acoustic
structure of the CMB~\citep{Planck:2018}. The second is that this invariant
pattern should be interpreted as evidence for a specific microscopic
ontology: cold dark matter particles, warm dark matter, self-interacting dark
matter, or something else entirely~\citep{Milgrom:1983,Verlinde:2017,Martens:2022,Kosso:2013}.
ISR is realist about the former and agnostic about the latter. The pattern is
invariant across admissible representations; the ontology used to encode it
is not. That is not a retreat from realism, but an attempt to place realism
where the epistemic warrant is strongest.

This operational criterion also supplies ISR's main response to Newman's
objection \citep{Newman:1928}. Newman showed that a purely structural
description, if construed only extensionally as a set of abstract relations
over an unspecified domain, risks collapsing into a trivial claim about
cardinality. ISR avoids this result because its invariants are neither bare
formal relations nor unconstrained structural placeholders. They are
\emph{empirically anchored}: tied to concrete quantitative constraints such
as observed power spectra, Fisher eigendirections, and information bounds.
They are also \emph{intensionally specified}: what makes them invariants is
not merely that they instantiate some abstract relation, but that they play a
determinate role in constraining inference across representations. The
acoustic structure of the CMB is not just ``some relation'' among temperature
anisotropies; it is a specific, quantitatively rigid pattern whose
preservation constrains any empirically adequate model. That empirical
rigidity is what deprives the Newman problem of its force in the present
context. Section~\ref{sec:conclusion} returns briefly to this point after the
formal discussion of \S\ref{sec:info_geom}.

\subsubsection{The embedded observer}
\label{sec:embedded}

The third distinction concerns the status of the observer. Standard
structural realism, whether epistemic or ontic, tends to characterize
structure from something close to a view from nowhere~\citep{Nagel:1986}: the
relations that constitute reality are those encoded in the formalism,
specified independently of any particular epistemic or observational
standpoint. In cosmology, however, that idealization is difficult to sustain.

Observers are embedded within the very system they seek to understand. They
are confined to the interior of a past light cone, limited by causal
horizons, finite sky coverage, and finite instrumental
reach~\citep{Ellis:1975,Ellis:2014,Rindler:1956}. They are also constrained by
cosmic variance, which reflects the fact that only one realization of the
relevant stochastic processes is available to observation~\citep{Knox:1995,Tegmark:1997,Scott:2016}.
These are not merely local practical obstacles. They are structural features
of the epistemic situation in which cosmological inquiry takes place.

ISR takes this fact seriously. The informational geometry of model space is
not specified once and for all in abstraction from all observational
conditions; it is conditioned by the observing configuration. The Fisher
information matrix depends on sky fraction, instrumental noise, redshift
coverage, bandpass, foreground treatment, and other features of the actual
inferential setup~\citep{Tegmark:1997,Heavens:2009,Schaefer:2016}. What
counts as invariant, therefore, is invariant \emph{for embedded observers}
under admissible variations of representation, not for a hypothetical
external intelligence with unrestricted access to all degrees of freedom.

This does not collapse objectivity into subjectivity. On the contrary, it
yields a more realistic conception of objectivity for cosmology: objectivity
within a physically defined epistemic horizon. In this respect ISR shares
something with perspectival realism~\citep{Massimi:2018,Giere:2006} and with
relational currents in the interpretation of quantum
mechanics~\citep{Rovelli:1996,Healey:2017}. But it adds an important further
claim: the information-geometric framework gives explicit criteria for which
features survive variations in perspective and which do not. Invariance is
not whatever appears stable from one standpoint; it is what remains stable
across the relevant class of admissible standpoints and descriptions.

This embedded-observer picture also aligns naturally with developments in
holography and quantum gravity, where horizons function not merely as
geometric curiosities but as information-theoretic boundaries on accessible
degrees of freedom~\citep{Bousso:2002,Jacobson:1995,Bekenstein:1981}. If
spacetime geometry is itself emergent from deeper informational structure, as
some contemporary approaches suggest~\citep{tHooft:1993,Susskind:1995,Maldacena:1999,VanRaamsdonk:2010,Swingle:2012,Bousso:1999},
then a realism grounded in informational invariants is arguably better
positioned than one that anchors objectivity at the level of spacetime
geometry itself. The point here is not that ISR depends on any one program in
quantum gravity. It is that the observer-embedded character of cosmological
knowledge is not a temporary defect to be idealized away. It is part of what
any adequate philosophical account must explain.

\subsection{Relation to Constructive Empiricism}

ISR also requires careful distinction from constructive empiricism, with
which it shares a visible surface affinity. Van Fraassen's position is well
known: the aim of science is empirical adequacy, and acceptance of a theory
does not require belief in the reality of its unobservable
posits~\citep{vanFraassen:1980}. In one respect, ISR agrees. The empirical
success of a cosmological model does not by itself warrant confidence in the
full ontology that the model deploys.

The agreement ends there. ISR makes a stronger claim about what cosmological
inquiry secures. It holds that the informational invariants isolated in the
previous sections are objective features of the world as available to
embedded observers, not merely features of our descriptions or summaries of
what has been observed. The correlation structure of the CMB, the geometry of
distinguishability in model space, and the information bounds imposed by
causal accessibility are not conveniences of our theories. They are real
constraints that any adequate cosmological representation must respect.

The difference may be put simply. The constructive empiricist says:
successful cosmological models save the phenomena, and no more belief is
required. ISR says: successful cosmological models do more than save the
phenomena; they encode invariant informational structure that is itself a
proper object of realist commitment. That commitment is modest, because it
does not license unrestricted belief in the unobservable ontology of a
favored theory. But it is realist nonetheless, because it attributes
objective status to features that are not exhausted by the observable in van
Fraassen's sense and that remain stable across empirically adequate
representations.

\subsection{Between No Miracles and Pessimistic Meta-Induction}

A final contrast concerns the broader dialectic of realism itself. The
familiar no-miracles argument holds that the empirical success of science
would be miraculous if our theories did not capture something real about the
world~\citep{Putnam:1975,Psillos:1999}. Any position that qualifies
ontological realism must therefore explain scientific success without
dissolving it into coincidence or mere convenience.

ISR offers such an explanation. The success of a cosmological model such as
$\Lambda$CDM is not accidental because the model captures real structure. But
the relevant structure is not identified with the literal truth of the
model's internal ontology. What the model gets right, when it succeeds, is
the invariant informational structure accessible to observation: the correct
correlation patterns, the right distinguishability relations, and the
constraints imposed by informational bounds. The model succeeds because it
encodes that structure efficiently, not because it has thereby uniquely
revealed the universe's final ontology.

This preserves the central insight of the no-miracles argument while
softening its ontological conclusion. The miracle would be if a model
reproduced the empirical and inferential structure of cosmological
observation without latching onto anything real. ISR denies that this is what
happens. But it equally denies that the only non-miracle explanation is
literal commitment to one theory's ontology.

ISR also reframes the pessimistic meta-induction. Laudan's argument turns on
the historical fact that scientifically successful theories have often turned
out to be ontologically mistaken~\citep{Laudan:1981}. The cases discussed in
Section~\ref{sec:underdetermination}---ether, caloric, Newtonian absolute
space---give that argument much of its force. ISR accepts the cautionary
lesson. If history teaches anything, it is that ontology is revisable. But
ISR adds that not everything equally revises. Across theory change,
empirical success is not preserved at random. What successor theories inherit
are the structured constraints their predecessors had already captured: the
power spectra, distance relations, and stable statistical patterns that new
theories must reproduce if they are to count as improvements at all. In that
sense, ISR turns the pessimistic meta-induction from a threat to realism into
a reason to relocate realist commitment. What survives theory change is
neither the whole ontology nor nothing at all, but the informational
invariants that constrain any successor representation.

\section{The Information Geometry of Cosmological Inference}
\label{sec:info_geom}

Information geometry provides an operational criterion for ISR’s notion of invariance.
If the objective content of cosmology lies in features preserved across empirically
adequate representations, then we require a framework that identifies which features
are invariant under changes of parameterization, model formulation, and inference
pipeline. Information geometry supplies such a framework by characterizing statistical
models in terms of their distinguishability properties.

The central idea is straightforward. A cosmological model does not merely assign values
to parameters; it defines a probability distribution over observable data. The geometry
of the space of such distributions encodes which models can be distinguished, which are
degenerate, and how observational precision partitions theory space. These geometric
features are not artifacts of a particular parameterization or ontology. They are
properties of the relationship between models and data. In this sense, they provide a
natural realization of the informational invariants identified in the previous section.

\subsection{Fisher Geometry and Representation-Independence}

Let $\mathcal{M}$ be a family of cosmological models parameterized by
$\boldsymbol{\theta}$, with likelihood $p(\mathbf{d}\,|\,\boldsymbol{\theta})$.
The Fisher information matrix
\begin{equation}
\mathcal{F}_{ij}(\boldsymbol{\theta}) =
\left\langle
\frac{\partial \ln p}{\partial \theta^i}
\frac{\partial \ln p}{\partial \theta^j}
\right\rangle
\end{equation}
defines a Riemannian metric on $\mathcal{M}$, the Fisher--Rao metric
\citep{Rao1945,Amari:2000,Cencov:1982}.

Its interpretive significance is direct: the infinitesimal line element
\begin{equation}
ds^2 = \mathcal{F}_{ij}\, d\theta^i d\theta^j
\end{equation}
measures statistical distinguishability. Nearby models with small $ds^2$ are
observationally indistinguishable; large distances correspond to models that
make detectably different predictions.

Two properties make this structure central to ISR.

First, \emph{reparameterization invariance}. The Fisher metric transforms covariantly
under smooth changes of coordinates, so its geometric content does not depend on how
model space is parameterized. This directly supports ISR’s claim that epistemically stable content 
should not depend on representational choices.

Second, \emph{uniqueness}. By \v{C}encov’s theorem, the Fisher--Rao metric is the
unique Riemannian metric invariant under sufficient statistics. It is therefore not an
arbitrary geometric overlay, but the canonical way of encoding the information content
of a statistical model.

From the ISR perspective, this matters because it identifies a concrete class of
representation-independent structures: distances, eigenvalues, and degeneracy
directions in model space. These are invariant under relabeling and therefore
constitute candidates for the shared empirical structure across models.

\subsection{Distinguishability and Degeneracy Structure}

The most important ISR-relevant feature of Fisher geometry is the structure of
\emph{degeneracies}. Directions $\delta\boldsymbol{\theta}$ satisfying
\begin{equation}
\mathcal{F}_{ij}\, \delta\theta^j \approx 0
\end{equation}
correspond to changes in model parameters that leave observational predictions
essentially unchanged.

In standard practice, such degeneracies are treated as nuisances. From the ISR
perspective, they are \emph{invariants}. A degeneracy direction is not a flaw in a
parameterization; it is a structural fact about what the data cannot distinguish.

The eigenstructure of $\mathcal{F}_{ij}$ makes this explicit. Large eigenvalues
identify well-constrained combinations of parameters (“stiff” directions), while
small eigenvalues identify poorly constrained or prior-sensitive combinations
(“sloppy” directions) \citep{Tegmark:1997,Huterer:2003,Transtrum:2015}.

This leads to a crucial shift in interpretation. Cosmological data do not primarily
determine individual parameters; they determine combinations of parameters defined by
the data’s sensitivity structure. The eigenvalue spectrum---not the named
parameters---captures what is invariant.

\subsection{Illustration: CMB Constraints}

This point can be seen clearly in the cosmic microwave background. The standard
$\Lambda$CDM parameter set
\begin{equation}
(\Omega_b h^2,\; \Omega_c h^2,\; H_0,\; \tau,\; n_s,\; A_s)
\end{equation}
is not directly measured parameter by parameter. Instead, the CMB constrains
specific combinations through its sensitivity to acoustic peak structure
\citep{Planck:2018,Ma:1995,Seljak:1996}.

For example:
- The combination $A_s e^{-2\tau}$ is tightly constrained, while $A_s$ and $\tau$
  separately are not.
- Geometric degeneracies link curvature, dark energy, and $H_0$
  \citep{Efstathiou:1999,Bond:1997}.

From a standard realist perspective, parameter estimates are interpreted as
measurements of physical quantities. ISR offers a different reading. What is
objectively fixed is the \emph{distinguishability structure}: which combinations of
features the data constrain and which they do not.

The parameters themselves are coordinates. The eigenstructure is the invariant.

\subsection{KL Divergence and Equivalence Classes}
\label{sec:kl}

The Fisher metric captures local distinguishability. To extend the framework across
finite separations and across model families, one uses the Kullback--Leibler (KL)
divergence:
\begin{equation}
D_{\mathrm{KL}}(p \,\|\, q) =
\int p(\mathbf{d}) \ln \frac{p(\mathbf{d})}{q(\mathbf{d})} d\mathbf{d}
\end{equation}
\citep{Kullback:1951,Cover:1991}.

KL divergence measures the empirical difference between models at the level of data.
Two models are empirically equivalent if their KL divergence falls below the
detectability threshold set by the observational configuration.

This provides a precise realization of ISR’s equivalence classes. Models belong to
the same class when
\begin{equation}
D_{\mathrm{KL}} \lesssim \Delta D_{\mathrm{KL}}^{\mathrm{obs}}\,.
\end{equation}

Three features make KL divergence especially important for ISR:

\begin{itemize}
\item It is defined purely at the level of predicted data distributions, independent
of ontology or formalism.
\item It unifies local (Fisher) and global notions of distinguishability.
\item It provides a quantitative criterion for when models are empirically
indistinguishable.
\end{itemize}

In this way, KL divergence formalizes the central ISR claim: what matters for
objectivity is not the internal structure of a model, but what cannot be distinguished
at the level of observational predictions.

\subsection{Information Bounds and Observational Limits}

A third class of invariants concerns limits on accessible information. These arise
from both observational constraints and fundamental physical principles.

In the CMB, the number of measurable modes is finite \citep{Knox:1995,Tegmark:1997}:
\begin{equation}
I_{\mathrm{total}} \lesssim \sum_\ell (2\ell+1) f_{\mathrm{sky}}.
\end{equation}
This sets an upper bound on the information available
for parameter inference.

More fundamentally, entropy bounds---such as the Bekenstein and covariant entropy
bounds---limit the information content accessible within bounded regions of spacetime
\citep{Bekenstein:1981,Bousso:1999,Bousso:2002}. Cosmic variance imposes an
irreducible floor on uncertainty \citep{Scott:2016}.

These limits are independent of model choice. They constrain all admissible
representations and therefore qualify as informational invariants in the ISR sense.
They define not what has been measured, but what \emph{can be measured in principle}.


\subsection{Bayesian Inference and Prior-Dependence}
\label{sec:bayesian}

The formal development above has relied on the Fisher information 
matrix and the Kullback--Leibler divergence, both of which belong 
to the frequentist and information-theoretic traditions.  Modern 
cosmological inference, however, is overwhelmingly Bayesian: 
parameter constraints are reported as posterior distributions 
$p(\boldsymbol{\theta} \,|\, \mathbf{d})$ obtained by combining a 
likelihood $p(\mathbf{d} \,|\, \boldsymbol{\theta})$ with a prior 
$\pi(\boldsymbol{\theta})$ via Bayes' 
theorem~\citep{Trotta:2008,Hobson:2010}.  The Fisher matrix 
enters this framework as a Gaussian approximation to the 
posterior valid near the maximum-likelihood point, and the 
Cram\'er--Rao bound as an idealized limit on achievable 
precision~\citep{Cramer:1946,Rao1945}.  A natural objection 
arises: if cosmological inference is Bayesian, and if the 
posterior depends on the prior, then the ``informational 
structure'' identified in the preceding sections may be partially 
prior-dependent and therefore not invariant in the sense 
required.

This objection is well taken as a refinement, but it strengthens 
rather than weakens the ISR position.  Prior choice is itself a 
representational decision.  The selection of flat priors on 
$\Omega_c h^2$ versus flat priors on 
$\log \Omega_c h^2$, or the adoption of informative versus 
uninformative priors on~$\tau$, reflects modeling assumptions 
rather than features of the 
data~\citep{BergerBernardoSun:2009,Jeffreys:1961}.  In exactly 
the same way that ISR treats the decomposition of eigenstructure 
into named parameters as representation-dependent, it treats prior-dependent features of 
the posterior as representation-dependent.  What is invariant is 
what survives under reasonable prior variation---a criterion that 
connects naturally to the Bayesian literature on reference 
priors and prior 
sensitivity~\citep{BergerBernardoSun:2009,Kass:1996}.

This criterion can be made precise.  Features of the posterior 
that are robust under changes of prior---such as the location of 
well-constrained modes, the rank ordering of parameter 
sensitivities, and the existence of strong degeneracies---reflect 
the dominance of the likelihood over the prior in the relevant 
directions.  These are precisely the directions along which the 
Fisher information is large, the ``stiff eigenmodes''.  Conversely, features that are 
sensitive to prior choice---such as upper limits on poorly 
constrained parameters, the shape of the posterior in sloppy 
directions, or one-sided bounds on quantities like 
$\sum m_\nu$~\citep{Planck:2018,DESI:2025a}---are directions 
along which the data contribute little information and the prior 
does most of the work.  ISR classifies the former as invariant 
and the latter as representation-dependent, which is exactly the 
classification a Bayesian epistemologist would endorse: 
conclusions driven by the data are robust; conclusions driven by 
the prior are 
not~\citep{BergerBernardoSun:2009,Handley:2019,Efstathiou:2008}.

The convergence between the two frameworks can be stated sharply.  
Let $\boldsymbol{e}_\alpha$ denote the eigenvectors of the Fisher 
matrix with eigenvalues 
$\lambda_\alpha$.  In the direction~$\boldsymbol{e}_\alpha$, the 
posterior width is approximately
\begin{equation}
\sigma_\alpha^2 \;\approx\; 
\frac{1}{\lambda_\alpha + \lambda_\alpha^{(\pi)}}\,,
\end{equation}
where $\lambda_\alpha^{(\pi)}$ characterizes the curvature of the 
log-prior in the same 
direction~\citep{Tegmark:1997,Heavens:2009}.  When 
$\lambda_\alpha \gg \lambda_\alpha^{(\pi)}$, the posterior is 
dominated by the likelihood and is prior-insensitive---this is a 
stiff direction.  When 
$\lambda_\alpha \ll \lambda_\alpha^{(\pi)}$, the posterior 
reflects the prior rather than the data---this is a sloppy 
direction.  The stiff/sloppy classification  thus maps directly onto the 
prior-sensitive/prior-insensitive classification of Bayesian 
analysis.  The informational invariants of ISR are precisely those 
features of cosmological inference that are robust under 
\emph{both} reparameterization (as shown by the Fisher metric) 
\emph{and} prior variation (as shown by Bayesian sensitivity 
analysis).

The Bayesian framework therefore provides a second, independent 
route to the same invariants that information geometry identifies 
through eigenvalues and degeneracies.  Far from undermining the 
ISR framework, the Bayesian perspective supplies a complementary 
lens through which the same structures can be recognized---and a 
further argument for their objectivity.  A feature that is 
invariant under changes of parameterization, stable under 
variation of priors, and robust across independent 
datasets~\citep{Handley:2019,Raveri:2019,Lemos:2021} has a 
strong claim to reflect the informational structure of the data 
rather than the representational choices of the analyst.

\subsection{Cosmological Tensions as Informational Mismatches}
\label{sec:tensions}

The information-geometric framework developed above provides a 
precise language for analyzing a phenomenon of growing importance 
in contemporary cosmology: persistent discrepancies between 
parameter values inferred from independent observational probes.  
Rather than treating these tensions as straightforward conflicts 
between measurements of the same quantity, ISR reframes them as 
structural features of the informational landscape---mismatches 
between the constraints that distinct inference pipelines impose 
on model space.

\subsubsection{The Hubble tension}

The most prominent case is the Hubble tension: the $\sim\!5\sigma$ 
discrepancy between the value of~$H_0$ inferred from CMB data 
under $\Lambda$CDM~\citep{Planck:2018} and that obtained from 
local distance-ladder 
measurements~\citep{Riess:2022,Freedman:2021}.  This tension is 
routinely presented as a conflict between two measurements of the 
same physical quantity.  In the information-geometric framework, 
the situation admits a more careful description.

The early-universe inference of $H_0$ is not a direct 
measurement but a model-dependent projection: a specific linear 
combination of $\Lambda$CDM parameters, constrained along a stiff 
eigendirection of the CMB Fisher matrix.  The local measurement, by 
contrast, is the output of a different inference pipeline---one 
involving Cepheid calibration, Type~Ia supernova standardization, 
and tip-of-the-red-giant-branch photometry---with its own 
systematic error budget and its own implicit model 
assumptions~\citep{Riess:2022,Freedman:2021,Efstathiou:2021}.  
What is directly given in each case is not a value of $H_0$ as 
such, but a set of observational data processed through a 
structured chain of modeling choices.

In the language of \S\ref{sec:kl}, the tension indicates that 
the joint KL divergence between the data distributions predicted 
by a single $\Lambda$CDM model and the distributions actually 
observed by the two probes exceeds the detectability 
threshold---the two datasets are informationally inconsistent 
under the assumption that both are generated by the same point in 
$\Lambda$CDM parameter 
space~\citep{Handley:2019,Raveri:2019,Lemos:2021}.  Whether 
resolution requires modifying the cosmological model (by 
introducing early dark energy, additional relativistic species, 
or non-standard recombination 
physics~\citep{DiValentino:2021,Schoeneberg:2022,Poulin:2019}), 
revising astrophysical systematics, or reinterpreting the 
inference procedure remains an open question within the 
informational landscape.

What matters for ISR is the structural character of this 
situation.  The tension does not straightforwardly falsify a 
single hypothesis; it reveals a mismatch within the 
representational network through which cosmological knowledge is 
mediated.  The invariant content is the structured inconsistency 
itself---the fact that the joint informational constraints 
imposed by early- and late-time observations resist 
accommodation within a single equivalence class---rather than 
any particular diagnosis of its origin.

\subsubsection{The $S_8$ tension and pipeline dependence}

A second cosmological tension sharpens this analysis.  The $S_8$ 
parameter, which quantifies the amplitude of matter fluctuations 
on scales of $8\,h^{-1}\,\mathrm{Mpc}$, has been the subject of 
a persistent discrepancy between CMB-derived predictions and 
measurements from weak gravitational lensing, galaxy clustering, 
and cluster abundances at lower 
redshifts~\citep{Heymans:2021,DES:2022,Amon:2023}.  In the 
eigenstructure language, $S_8$ is 
a derived quantity---a specific projection of the posterior onto 
a direction in parameter space that combines $\Omega_m$ and 
$\sigma_8$---whose constraining power and degeneracy structure 
differ between the CMB and weak-lensing Fisher matrices.

The evolving status of the $S_8$ tension is particularly 
instructive.  Recent analyses, including the KiDS Legacy 
release, have found the tension reduced relative to earlier 
estimates---a shift attributed in part to improvements in the 
analysis pipeline itself, including revised treatments of 
intrinsic alignments, photometric redshift calibration, and 
baryonic feedback 
modeling~\citep{Wright:2025,Li:2023,Asgari:2021}.  The 
significance of the ``tension'' changed not because of new data 
about the universe but because of refinements in how existing 
data are processed.  This is precisely the pattern ISR predicts: 
the informational invariant---the joint constraint surface 
defined by CMB and weak-lensing data in the space of predicted 
$C_\ell$'s---is stable, while the particular numerical value of 
$S_8$ extracted within a given framework is partially 
representation-dependent.

\subsubsection{DESI and evolving dark energy}

A third case, rapidly developing, reinforces the general 
pattern.  The second data release of the Dark Energy 
Spectroscopic Instrument reports a preference for a 
time-varying dark energy equation of state at a significance of 
$2.8\sigma$ to $4.2\sigma$, depending on the supernova dataset 
included in the 
combination~\citep{DESI:2025a,DESI:2025b}.  The cosmological 
constant ($w = -1$, constant in time) lies outside the preferred 
region across all dataset combinations, though the statistical 
weight of the evidence varies with the choice of 
catalog~\citep{DESI:2025a}.

From the ISR perspective, this development illustrates the 
invariant/representation-dependent distinction in real time.  
The informational invariant---the specific shape of the 
distance--redshift relation as jointly constrained by BAO, CMB, 
and supernovae---is robustly established.  What is under active 
revision is the encoding: whether this invariant is best 
captured by a cosmological constant, a $w_0 w_a$ 
parameterization~\citep{Chevallier:2001,Linder:2003}, a 
dynamical scalar field, or something else entirely.  The fact 
that the statistical significance of the departure from 
$\Lambda$ depends sensitively on which supernova catalog is 
used (Pantheon+, Union3, DESY5) further underscores that the 
``result'' is a joint product of data and inference pipeline, 
not a clean readout from 
nature~\citep{DESI:2025a,Rubin:2025}.  ISR predicts exactly 
this pattern: the invariant holds while the encoding is revised.

The information-geometric framework developed in this section 
provides the formal backbone for the philosophical claims 
advanced in \S\ref{sec:ISR}: it identifies a concrete class of 
reparameterization-invariant structures, makes the distinction 
between invariant and representation-dependent features 
operationally sharp, and gives precise content to the notion of 
cosmological models as alternative encodings of a common 
informational structure.  It is important, however, to be 
explicit about what the formalism does \emph{not} settle.

The Fisher--Rao metric is a local object: it characterizes 
distinguishability in the neighborhood of a point in parameter 
space, not the global topology of theory space.  It presupposes 
that models can be parameterized smoothly, which may fail for 
theories with different numbers of parameters or fundamentally 
different mathematical structures.  The KL divergence extends the 
framework to finite separations, but it still requires that two 
models generate distributions over a common data space---a 
condition that is well satisfied within cosmological practice but 
cannot be taken for granted in comparisons across deeply 
different theoretical frameworks.  More fundamentally, the 
information-geometric framework identifies which features of 
cosmological inference are representation-independent; it does 
not, by itself, determine whether those features should be 
interpreted realistically.  The step from ``invariant under 
reparameterization'' to ``objectively real'' is a philosophical 
inference, not a mathematical theorem, and it carries commitments 
(spelled out in \S\ref{sec:ISR}) that go beyond what the 
formalism alone can justify.  ISR's claim is that the 
information-geometric criterion provides the \emph{best 
available} basis for such inferences in cosmology, not that it 
renders them automatic.


\subsection{What the Formalism Does \emph{Not} Show}

It is essential to be clear about the limits of this framework.

First, the Fisher metric is local. It characterizes distinguishability near a point in
model space and does not capture global structure by itself.

Second, both Fisher and KL frameworks assume that models define probability
distributions over a shared data space. This is well satisfied in cosmology but may
fail in more radical theory changes.

Third, and most importantly, the formalism does not by itself justify realism. The
fact that a structure is invariant under reparameterization does not entail that it is
ontologically fundamental. The step from invariance to realism is philosophical, not
mathematical.

ISR’s claim is therefore modest. Information geometry does not prove that
informational invariants are real. It shows that they are the best candidates for
objective content available within cosmological inference.

\subsection{Implications}

The information-geometric framework sharpens the ISR proposal in three ways.

First, it identifies a concrete class of representation-independent structures:
distinguishability geometry, degeneracy structure, and information bounds.

Second, it explains why named parameters and theory-specific ontologies are less
secure: they are coordinate-dependent projections of invariant structures.

Third, it provides a precise account of scientific progress. As observational
precision improves, equivalence classes defined by KL divergence split, and formerly
degenerate directions become distinguishable. Progress is therefore best understood
as refinement of informational structure rather than convergence to a uniquely fixed
ontology.

This completes the operational grounding of informational structural realism. The
formalism does not replace philosophical argument, but it shows that the notion of
informational invariance can be made precise, non-trivial, and directly connected to
the practice of cosmological inference.

\section{Objections and Limits}
\label{sec:objections}

The account developed in this paper aims to locate the inferentially accessible structure of
cosmology in cross-representational informational invariants. Given its
departure from both standard realism and empiricism, several natural
objections arise. This section addresses four of the most pressing concerns
and clarifies the limits of the proposal.

\subsection{Is ISR Just Empiricism in Disguise?}

A first worry is that informational structural realism collapses into a
sophisticated form of empiricism. If ISR declines to identify objective
content with theory-internal ontology, and instead emphasizes what is
preserved across empirically well-supported representations, does it not simply
restate the empiricist claim that only what is tied to observable phenomena
is epistemically warranted?

The resemblance is real but incomplete. Constructive empiricism maintains
that acceptance of a theory involves belief only in its empirical adequacy
with respect to observables~\citep{vanFraassen:1980}. ISR goes further. The
informational invariants it identifies are not restricted to directly
observable quantities, nor are they exhausted by the requirement that a model
``save the phenomena.'' They include higher-order features of the inferential
structure linking models to data: correlation patterns, distinguishability
relations, and bounds on accessible information.

These features are realist in a specific and limited sense. They are
\emph{cross-representational}: any empirically adequate description must
preserve them. They are also \emph{constraint-based}: they encode what the
world permits and forbids in observational structure, rather than merely
recording what has been observed. For example, the degeneracy structure of
cosmological inference---the existence of directions in model space along
which predictions are indistinguishable---is not itself an observable, but it
is a stable feature of the relationship between models and data. Likewise,
information bounds and cosmic variance are not observables, but they are
objective constraints on what can be known.

ISR therefore occupies an intermediate position. It agrees with empiricism
that ontology is underdetermined by empirical success, but it rejects the
restriction of epistemic commitment to observables alone. The invariants it
identifies are not merely epistemic conveniences; they are the most stable
features of how the world is accessible to observation. In that sense, ISR is
not empiricism in disguise, but a constrained form of realism about
informational structure.

\subsection{Observer-Relativity and Objectivity}

A second objection concerns observer-relativity. Informational invariants are
defined relative to an observational configuration: sky coverage, noise
properties, instrumental sensitivity, and causal accessibility all affect
which features of model space are distinguishable. If invariants depend on
such factors, does objectivity collapse into perspectival or even subjective
dependence?

The force of this objection depends on a strong assumption: that objective
features must be independent of all observational conditions. ISR rejects
this assumption. In cosmology, observers are physically embedded within the
system they study, constrained by causal horizons, finite information
channels, and statistical limitations such as cosmic variance
\citep{Ellis:1975,Scott:2016}. Any account of objectivity that ignores these
constraints risks describing an idealized perspective unavailable to any
actual or possible observer.

ISR therefore adopts a notion of \emph{domain-relative objectivity}. An
informational invariant is objective if it is stable across admissible
representations within a given observational domain. That domain includes not
only the phenomena under study but also the physical constraints on access to
them. This is not subjectivity. Different observers operating under the same
observational conditions will recover the same invariants. What varies is not
the truth of the invariant, but the partition of model space induced by
changes in observational precision.

This situation has close analogues elsewhere in physics. Thermodynamic
quantities such as entropy are defined relative to coarse-grainings, yet are
not thereby subjective. Likewise, in cosmology, the invariants identified by
ISR are indexed to the informational structure available to embedded
observers, but remain stable across all admissible representations of that
structure. Domain-relativity, in this sense, is compatible with objectivity.

\subsection{Is Admissibility Too Permissive?}

A third concern is that ISR’s notion of admissibility may be too permissive.
If informational invariants are defined across a broad class of
representations, what prevents arbitrary or pathological models from being
included? Without sufficiently strong constraints, the equivalence-class
framework risks triviality: everything could be equivalent to everything
else.

ISR addresses this by imposing minimal but substantive conditions on
admissible representations. A model is admissible only if it (i) defines a
probability distribution over a shared data space, (ii) yields a
well-posed likelihood structure, and (iii) is not already ruled out by
existing observations at the relevant level of precision. These constraints
are weak enough to allow representational plurality, but strong enough to
exclude arbitrary constructions.

The requirement of a shared data space is particularly important. Two models
can be compared informationally only if they make predictions about the same
observables---for example, the CMB power spectrum, galaxy clustering
statistics, or distance--redshift relations. Likewise, the requirement of a
likelihood structure ensures that models can be embedded within a common
inferential framework. Finally, empirical viability restricts attention to
representations that genuinely compete to account for the data.

These constraints align with standard scientific practice. Cosmologists do
not treat arbitrary mathematical constructions as serious alternatives; they
consider models that are predictive, mathematically coherent, and empirically
testable. ISR formalizes this practice without over-constraining it. The
result is a notion of admissibility that is neither trivial nor excessively
restrictive.

\subsection{Limits of the Formalism}

A final objection concerns the reliance on information geometry. The Fisher
metric is local, the KL divergence depends on the choice of likelihood, and
both frameworks presuppose a common data space. Do these limitations
undermine the claim that informational invariants provide a robust basis for
realism?

The short answer is that they do not, but they do delimit the scope of the
proposal. Information geometry is not intended to provide a complete
foundational account of cosmology. It is a tool for identifying
representation-independent features of the relationship between models and
data. Its limitations are therefore expected.

First, the Fisher--Rao metric captures local distinguishability and does not
fully characterize global structure in model space. This is why it must be
supplemented by measures such as the KL divergence for finite separations.
Second, both Fisher and KL frameworks depend on the specification of a
likelihood function, which is itself part of the modeling process. Third,
comparisons across radically different theoretical frameworks may strain the
assumption of a shared data space.

ISR accommodates these limitations by treating information geometry as an
\emph{operational criterion}, not a foundational theorem. The claim is not
that Fisher geometry uniquely determines what is real, but that it provides
the best available method for identifying invariants within the domain of
cosmological inference. Where its assumptions break down, the framework may
require extension or modification.

A related concern is that the KL divergence and Bayesian inference introduce
prior dependence. ISR’s response is that prior-sensitive features are, by
definition, not invariant. The framework explicitly distinguishes between
structures that are robust under changes of prior and those that are not.
This distinction is not a weakness but a diagnostic: it separates
representation-dependent features from those that carry what cosmology most securely establishes.

The broader point is that ISR is a \emph{framework}, not a theorem. It does
not claim to derive realism from formal considerations alone. Rather, it
argues that informational invariance provides the most defensible basis for
realist commitment given the epistemic situation of cosmology. The formalism
clarifies and supports this claim, but does not exhaust it.

\section{Conclusion: What Cosmology Tells Us}
\label{sec:conclusion}

This paper has argued for a more cautious interpretation of cosmological
success. The empirical achievements of $\Lambda$CDM and related frameworks
are undeniable, but they do not by themselves fix a uniquely privileged
ontology, nor even a uniquely privileged mathematical representation. What
cosmology secures most robustly, I have argued, is instead a class of
\emph{informational invariants}: features of observational structure that
persist across representations consistent with the data. These include
correlation structure, structures of statistical distinguishability, and
limits on accessible information. 

This proposal is realist, but selectively so. It agrees with more standard
forms of realism that cosmological success is not accidental, and it agrees
with structural realism that what survives theory change is more secure than
the transient ontology of any one theory. But it departs from both standard
realism and ontic structural realism in refusing to identify the structure of
the world with the mathematical content of a single favored formalism. What
matters, on the view defended here, is not structure within one theory, but
structure preserved across admissible representations. It also departs from
constructive empiricism by treating these invariants as objective features of
the world as available to embedded observers, rather than as mere summaries
of observable phenomena.

The role of information geometry in this argument has been deliberately
modest. The Fisher--Rao metric and the Kullback--Leibler divergence do not
settle the metaphysics of cosmology. They do, however, provide an operational
criterion for ISR's central distinction between invariant and
representation-dependent features. They show how distinguishability,
degeneracy structure, and equivalence classes can be characterized in a way
that does not depend on arbitrary parameterization. The formalism therefore
does not prove ISR, but it makes the view sharper, less metaphorical, and
more closely connected to cosmological practice.

One consequence is especially important. On the account developed here,
scientific progress in cosmology should not be understood primarily as
convergence toward a uniquely correct ontology. It is better understood as
the progressive refinement of informational structure. As observations
improve, equivalence classes split; formerly degenerate directions become
distinguishable; the geometry of model space is resolved more finely. What
grows is not necessarily our proximity to the final inventory of what exists,
but the precision with which the universe constrains admissible
representations of itself.

This picture helps clarify the status of some of cosmology's most prominent
ontological posits. In the case of dark matter, for example, the robust
achievement of cosmology is the establishment of a stable cross-probe
informational pattern---acoustic peak structure, clustering statistics,
lensing discrepancies, and related constraints---even if the interpretation
of that pattern in terms of a particular microscopic ontology remains open
\citep{Planck:2018,Rubin:1980,Clowe:2006,BOSS:2017}. The same general lesson
applies more broadly: cosmology often secures the invariant before it secures
the encoding.

The view defended here is intentionally modest about what cosmology tells us,
but not modest about its achievements. Cosmology does yield objective
knowledge. It does so under severe epistemic constraints, from a single
embedded vantage point, and through heavily theory-mediated forms of
observation and inference. Precisely for that reason, its most secure
achievements are not best identified with the ontology of any one successful
model. They are best identified with the informational structure that all
successful models must preserve.

Cosmology does not reveal what the universe is made of; it reveals the
structure of what can be known about it.

\bibliography{bib}

\end{document}